\documentclass[runningheads]{llncs}
\usepackage{amsmath}
\usepackage[T1]{fontenc}
\usepackage{graphicx}
\usepackage[T1]{fontenc}
\usepackage{graphicx}   
\usepackage{subcaption} 
\usepackage{booktabs} 
\usepackage{algorithm}
\usepackage{algpseudocode}
\usepackage{amssymb}  
\usepackage{listings}  
\usepackage{xcolor}  
\algrenewcommand\algorithmicindent{1.5em} 
\newcommand{\prc}{\textit{prc}} 
\newcommand{\tacc}{\tau_{\text{acc}}} 
\newcommand{\tfair}{\tau_{\text{fair}}} 
\usepackage[utf8]{inputenc}
\usepackage{pgfplots}
\usepackage[absolute,overlay]{textpos}
\pgfplotsset{compat=1.18}
\usepackage{adjustbox}
\usepackage{makecell}
\usepackage{xcolor} 
\usepackage{bm}
\usepackage{hyperref}
\hypersetup{
    colorlinks=true,
    linkcolor=blue,
    filecolor=magenta,      
    urlcolor=cyan,
}

\begin{document}

\title{Achieving Fair Skin Lesion Detection through Skin Tone Normalization and Channel Pruning}

\author{Zihan Wei and Tapabrata Chakraborti}

\institute{Dept of Medical Physics \& Biomedical Engineering, University College London, UK\\
\email{zczqeic@ucl.ac.uk, tchakraborty@turing.ac.uk, t.chakraborty@ucl.ac.uk}\\
}

\maketitle             

\begin{abstract}
Recent works have shown that deep learning based skin lesion image classification models trained on unbalanced dataset can exhibit bias toward protected demographic attributes such as race, age, and gender. Current bias mitigation methods usually either achieve high level of fairness with the degradation of accuracy, or only improve the model fairness on a single attribute. Additionally usually most bias mitigation strtageies are either pre hoc through data processing or post hoc through fairness evaluation, instead of being integrated into the model learning itself. To solve these existing drawbacks, we propose a new Individual Typology Angle (ITA) Loss-based skin tone normalization and data augmentation method that directly feeds into an adaptable meta learning-based joint channel pruning framework. In skin tone normalization, ITA is used to estimate skin tone type and adjust automatically to target tones for dataset balancing. In the joint channel pruning framework, two nested optimization loops are used to find critical channels. The inner optimization loop finds and prunes the local critical channels by weighted soft nearest neighbor loss, and the outer optimization loop updates the weight of each attribute using group wise variance loss on meta-set. Experiments conducted in the ISIC2019 dataset validate the effectiveness of our method in simultaneously improving the fairness of the model on multiple sensitive attributes without significant degradation of accuracy. Finally, although the pruning mechanism adds some computational cost during training phase, usually training is done offline. More importantly, the pruned network becomes smaller in size and hence has lower compute time at inference stage thus making it easier to deploy in low resource clinical settings.

\end{abstract}

\section{Introduction}
Skin cancer such as melanoma is one of the most prevalent cancers due to its high invasiveness and metastatic ability, and its incidence has increased in recent years~\cite{Ali2022}. Modern artificial intelligence (AI) powered methods have been shown to have strong potential for early detection of skin cancer from deromscopy and even phone images, but these methods still have few key drawbacks before they can be safely dpeloyed in practical clinical settings~\cite{ansari2024algorithmic}. Firstly, deep learning based AI models may lose fair generalisation across patient cohorts with respect to protected attributes like age, gender, race, thus making them less trustworthy in clinical situations~\cite{Kalb2023}. For example, caucasian populations are more susceptible towards skin lesions due to lower skin mealinin concentration. As a consequence, usually the usually available skin datasets heavily feature lighter skinned patients and hence any machine learning models picks up a training bias in favour of those groups, and hence do not perform well when presented lesions on darker skin~\cite{biaslessnas}. In high-stakes environments like healthcare, clinicians must be able to understand and justify AI-assisted decisions, especially when outcomes impact patient safety or involve legal liability~\cite{guidotti2018survey}. 

Various methods have been proposed to solve the class imbalance and skin tone bias problems in the existing skin lesion datasets. The pre-processing-based methods include massaging, reweighting, and sampling~\cite{kamiran2009preprocessing}. They focus on modifying (normalising or augmenting) the training data or instance weights before model training. While effective in some cases, pre-processing techniques can be limited by their model-agnostic nature and might not fully address bias arising during model training~\cite{Chen2024SkinCancer}. Other methods include reduction techniques that transform fair classification into a sequence of cost-sensitive classification problems~\cite{agarwal2018reductions}. These methods reformulate fairness constraints as part of the optimization objective but may require complex optimization procedures~\cite{Paxton2024MeasuringAIFairness}. Another line of work focuses on multi-task learning and domain adaptation techniques, such as Quasi-Pareto Improvement (QPI)~\cite{Yao2024}, which incorporates weighted loss functions and transfer learning to improve performance on minority subgroups. 

While these methods show promise, they still have significant drawbacks. Most of these methods still struggle with the accuracy/performance vs transparency/fairness tradeoff~\cite{bhattacharyya2024conformal}. Moreover, not enough studies have been done on mitigating multiple demographic biases simultaneously in lesion classification. Most existing approaches focus on single attributes such as skin tone or gender, lacking a comprehensive framework for addressing intersectional bias across multiple protected attributes~\cite{ren2024skincon}. This limitation gives rise to serious problems in machine learning fairness and health equity, as real-world clinical applications require models that perform equitably across diverse patient populations~\cite{chowdhury2021exploring}. By critically evaluating the strengths and limitations of the existing methods discussed above, we develop in this work an algorithmic fairness approach to solve the imbalanced classes and inherent multi-demographic bias in skin lesion classification simultaneously along with maintaining the performance accuracy, all integrated seamlessly in to an end-to-end multi-objective training regime. The contributions of this paper are as follows.

\begin{enumerate}
    \item We propose a novel approach utilizing an individual Typology Angle (ITA) Loss-based skin tone normalization and data augmentation method that has a rigorous mapping to the Fitzpatrick Skin Types (FST) utlising the CIELAB colour space.
    \item We introduce an adaptive blend training regime that changes the probability of each class being sampled for training according to the model performance on validation set. This novel strategy can make the model pay more attention on minor groups.
    \item We develop a new meta learning based joint channel pruning approach that assigns dynamic weights to each sensitive attribute in order to prune the channels. This strategy can simultaneously improve the alogrithmic fairness on several senstiive attributes like skin tone, age and gender during the iterative training process, thus also optimising accuracy at the same time. 
    \item There are 2 additional positive by-products of the channel pruning process. First, due to the channel pruning process we end up with a more optimised and lean design for the deep learning vision model. Second, again due to the channel pruning, a pot-hoc visualisation which channels were pruned vs which were not, superimposed onto the input images, provides us an additional layer of interpretability. 
\end{enumerate}

\section{Methods}
\begin{figure}
\includegraphics[width=\textwidth]{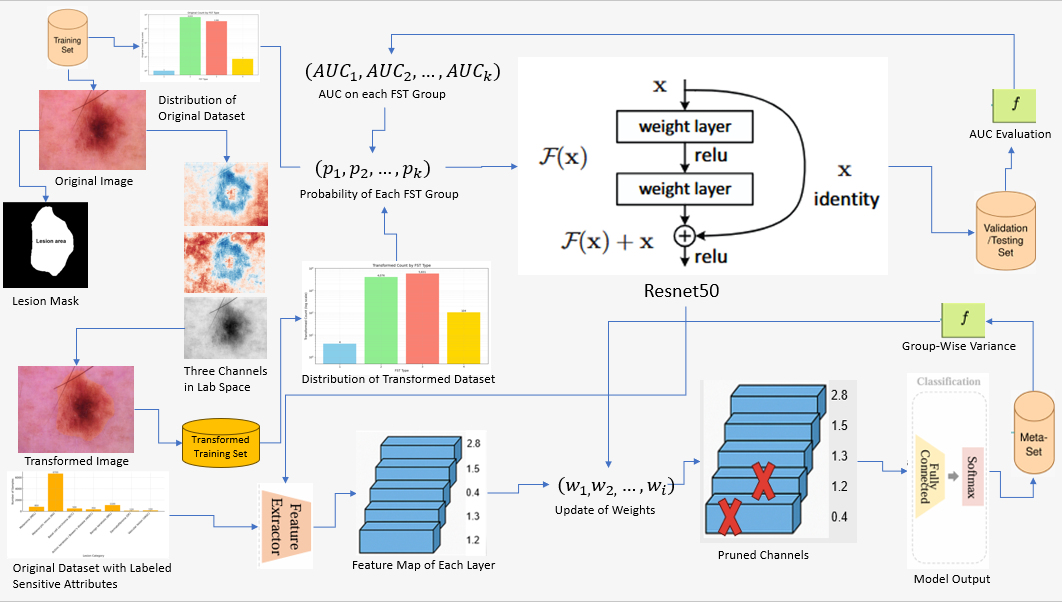}
\caption{Our Proposed Framework} \label{graphical abstract}
\end{figure}

This section introduces the public dataset used, the model architecture, training strategy, and the explainability aspect of learned features. The method is summarized in Figure~\ref{graphical abstract}.
\subsection{ISIC2019 Dataset}
\begin{figure}[htbp]
    \centering
    \includegraphics[width=\textwidth]{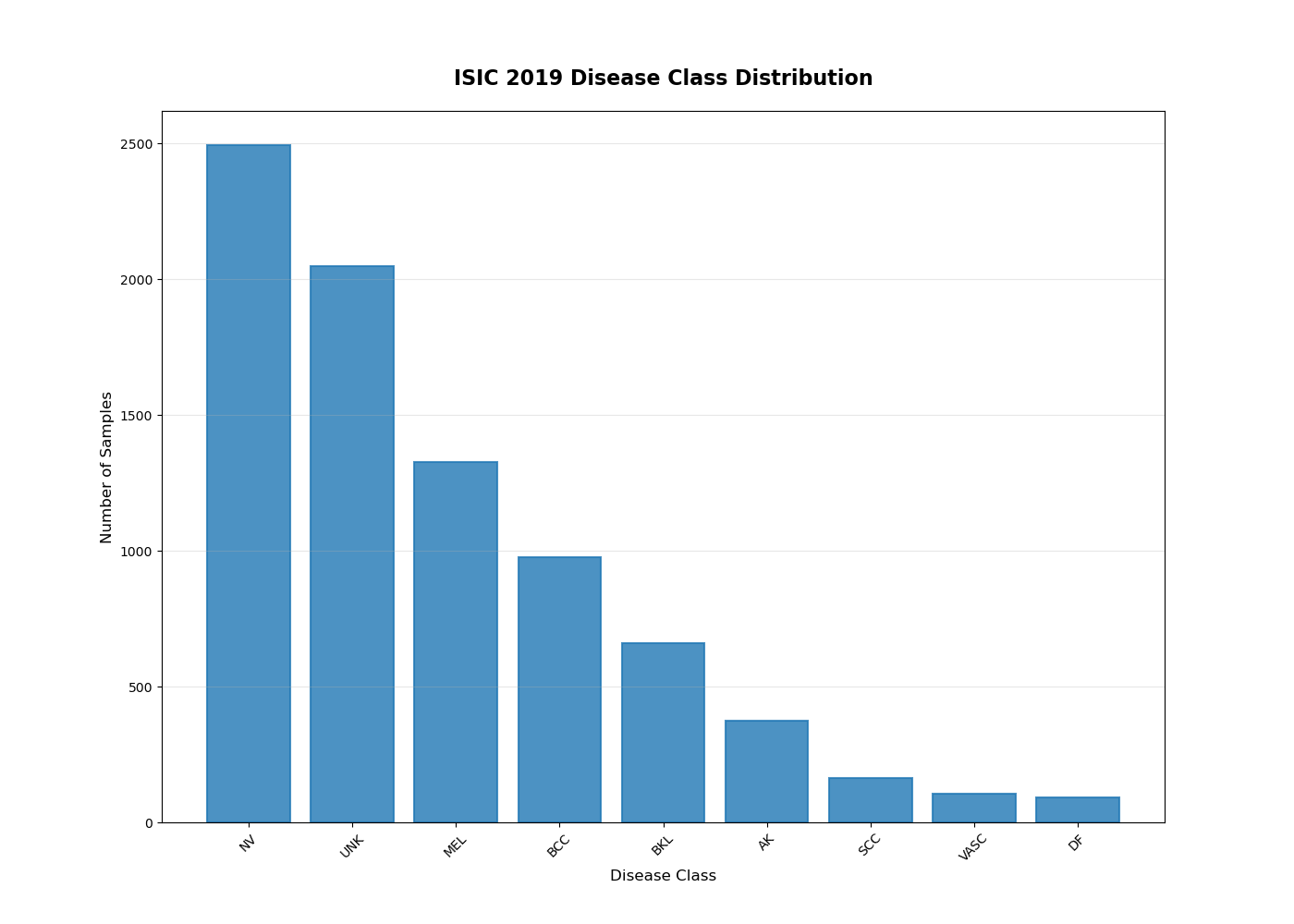} 
    \caption{Imbalance class distribution of lesions in ISIC2019 dataset.} 
    \label{fig:Lesion Distribution} 
\end{figure}

ISIC2019 (also known as HAM20000) is a large-scale publicly available dataset for skin lesion analysis, released as part of the International Skin Imaging Collaboration (ISIC) challenge~\cite{gessert2020skin}. It contains a total of 25,331 dermoscopic images and eight diagnostic labels, including melanoma, basal cell carcinoma, actinic keratosis, and benign nevi, among others. In addition to the images, metadata such as patient age, sex, and anatomical site is provided, which allows for fairness analysis across demographic subgroups. The dataset presents challenges such as class imbalance and demographic bias, making it a valuable benchmark for developing robust and fair diagnostic algorithms in dermatology. In this dataset, the number of patients with lighter skin tone is greater in number than darker skin tone, and that is consistent with higher prevalence of skin cancer among patients with less melanin concentration.
The distribution of lesion category is shown in Figure~\ref{fig:Lesion Distribution}. It can be seen that the distribution of lesion classes is also imbalanced. The number of Melanocytic nevus is much higher than others.

\subsection{Preprocessing: Skin Tone Normalization}
One common solution for unbalanced data problem is data augmentation. The idea of this technique is to generate new data by translating, rotating and transforming specific areas of original data in order to increase the amount of limited data. In skin lesion detection task, data augmentation is needed to balance the skin tone variation, which we achieve through the following steps. 

\subsubsection{Skin lesion masking}
\begin{figure}[htbp]
    \centering
    \begin{subfigure}[b]{0.48\textwidth}
        \centering
        \includegraphics[width=\textwidth]{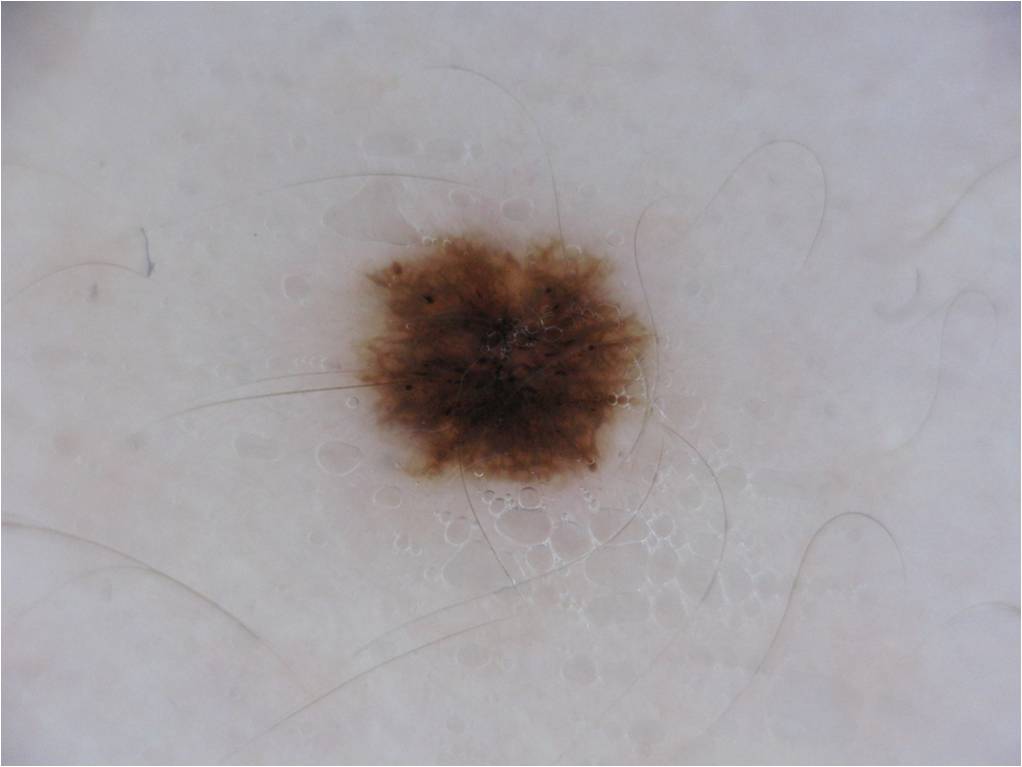}
        \caption{Original Image}
        \label{fig:image1a}
    \end{subfigure}
    \hfill
    \begin{subfigure}[b]{0.48\textwidth}
        \centering
        \includegraphics[width=\textwidth]{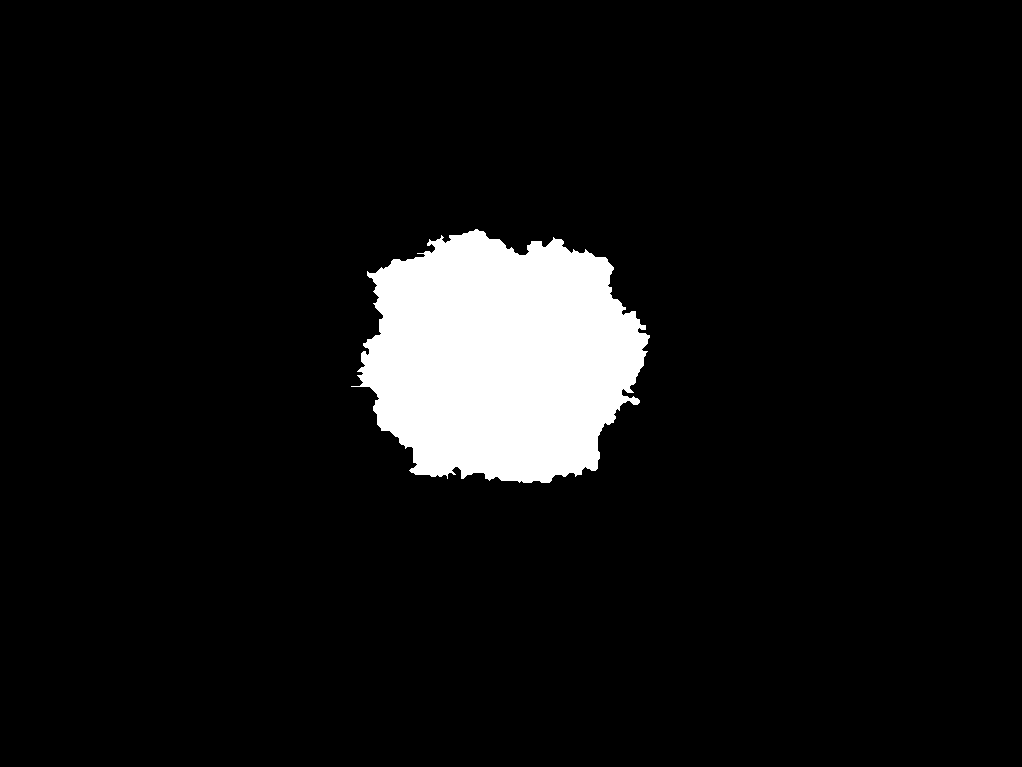}
        \caption{Mask}
        \label{fig:image1b}
    \end{subfigure}
    \caption{Segmentation results showing original image and its corresponding mask}
    \label{fig:segmentation}
\end{figure}

In skin tone augmentation, the lesion area should remain untouched during the transformation. Therefore, the skin lesion area needs to be masked before normalization. In this study, a standard U-net is trained using ISIC2018 dataset for masking~\cite{ronneberger2015unet}. The structure of U-Net consists of a contracting path (encoder) and a symmetric expanding path (decoder) with the learned feature speace at the junction. In this structure, downsampling operations increase the network's depth, which can help extract more abstract high-level features, while upsampling is used to restore the spatial dimensions of the image, enabling more precise localization. In our network, the encoder contains four double convolution operations inserted with batch normalization and activation function, and a max pooling layer at the end. This structure can reduce the spatial resolution and increase the feature channels to extract high level features. As a symmetrical module to encoder, the decoder takes the opposite operations to recover the spatial resolution. During encoding stage, a skip connection is used to join the feature images from encoder to corresponding layer in decoder, which can store the detailed information of high resolution and avoid gradient vanishing. This symmetrical network is easy to extend and it can integrate low-level and high-level features effectively. An example masking is shown in Figure~\ref{fig:segmentation}.

\subsubsection{Color Space Conversion}
The input image which is originally in RGB space is converted into CIELAB space before color transformation. CIELAB (also known as L*a*b*) is a device-independent color space. It was defined in 1976 by the International Commission on Illumination (Commission Internationale de l'Éclairage, abbreviated CIE) with the aim of establishing a model based on human visual perception to accurately and objectively measure and describe all colors perceived by the human eye~\cite{sharma2004ciede2000}. Like RGB space, CIELAB space also contains three channels but their functions are different. In CIELAB space, \textit{b} channel controls Lightness, \textit{a} channel controls Green and Red, and `B' channel controls Blue and Yellow. Among them, the variation in skin tone is relatively narrow along \textit{a} channel, so \textit{a} channel will not be considered in this task. CIELAB space has more advantages in skin tone transformation. RGB is a nonlinear color space which is closely tied to display devices and does not always align with human visual perception. Direct adjustments to RGB channels often introduce unnatural color shifts such as bluish or reddish tints, making it difficult to precisely control complex color attributes. Compared with RGB space, the numerical difference between two colors in CIELAB space is approximately proportional to the degree of difference perceived by the human eye, so it is more intuitive and allows more precise skin color adjustment. The conversion from RGB to CIELAB involves the following three main steps~\cite{plataniotis2000color}:

\paragraph{1. RGB Normalization and Linearization}
Given standard RGB values $R, G, B \in [0, 255]$, they are first normalized to the range $[0, 1]$:
\[
R' = \frac{R}{255}, \quad G' = \frac{G}{255}, \quad B' = \frac{B}{255}
\]
Next, gamma correction is applied to convert standard RGB to linear RGB:
\[
C_\text{linear} =
\begin{cases}
\frac{C'}{12.92}, & C' \leq 0.04045 \\
\left(\frac{C' + 0.055}{1.055}\right)^{2.4}, & C' > 0.04045
\end{cases}
\]
where $C' \in \{R', G', B'\}$ and $C_\text{linear} \in \{R_\text{lin}, G_\text{lin}, B_\text{lin}\}$.

\paragraph{2. Linear RGB to XYZ Color Space Conversion}
The linear RGB values are then converted to the CIE 1931 XYZ color space (using the D65 reference white point):
\[
\begin{bmatrix}
X \\
Y \\
Z \\
\end{bmatrix}
=
\begin{bmatrix}
0.4124 & 0.3576 & 0.1805 \\
0.2126 & 0.7152 & 0.0722 \\
0.0193 & 0.1192 & 0.9505 \\
\end{bmatrix}
\cdot
\begin{bmatrix}
R_\text{lin} \\
G_\text{lin} \\
B_\text{lin} \\
\end{bmatrix}
\]

\paragraph{3. XYZ to CIELAB Conversion}
Next, the XYZ values are normalized using the D65 reference white point:
\[
X_n = 95.047, \quad Y_n = 100.000, \quad Z_n = 108.883
\]
\[
x = \frac{X}{X_n}, \quad y = \frac{Y}{Y_n}, \quad z = \frac{Z}{Z_n}
\]

Define the function $f(t)$ as:
\[
f(t) =
\begin{cases}
t^{1/3}, & t > 0.008856 \\
7.787t + \frac{16}{116}, & t \leq 0.008856
\end{cases}
\]

Finally, the components of the CIELAB color space are computed as:
\[
L^* = 116 \cdot f(y) - 16
\]
\[
a^* = 500 \cdot (f(x) - f(y))
\]
\[
b^* = 200 \cdot (f(y) - f(z))
\]

\begin{figure}[htbp]
    \centering
    \begin{subfigure}[t]{0.32\textwidth}
        \centering
        \includegraphics[width=\linewidth, height=5cm, keepaspectratio]{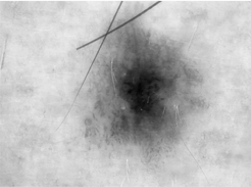}
        \caption{channel \textit{L}}
        \label{fig:imgL}
    \end{subfigure}
    \hfill
    \begin{subfigure}[t]{0.32\textwidth}
        \centering
        \includegraphics[width=\linewidth, height=5cm, keepaspectratio]{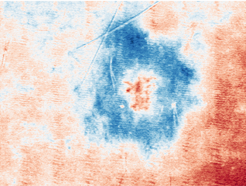}
        \caption{channel \textit{a}}
        \label{fig:imga}
    \end{subfigure}
    \hfill
    \begin{subfigure}[t]{0.32\textwidth}
        \centering
        \includegraphics[width=\linewidth, height=5cm, keepaspectratio]{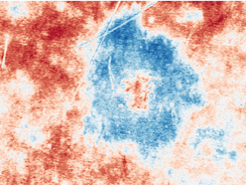}
        \caption{channel \textit{b}}
        \label{fig:imgb}
    \end{subfigure}
    \caption{Three channels in CIELAB Space}
    \label{fig:Three channels in CIELAB Space}
\end{figure}

The three channels in CIELAB space is shown in Figure~\ref{fig:Three channels in CIELAB Space}.

\subsubsection{Compute Individual Typology Angle}
\begin{figure}[htbp]
    \centering
    \includegraphics[width=0.8\textwidth]{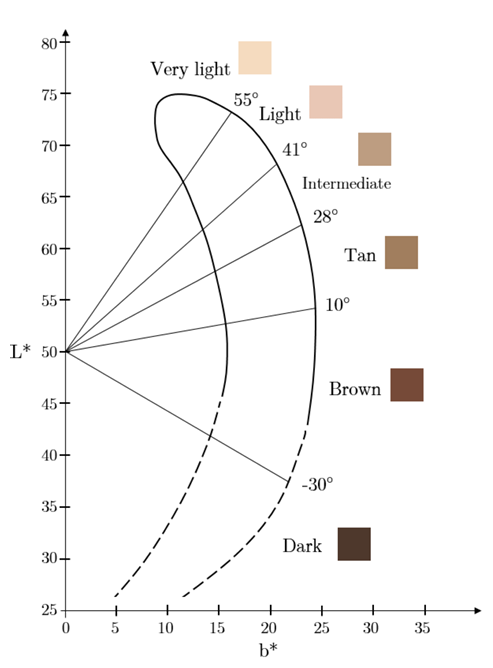} 
    \caption{Contributions of L and b to ITA value~\cite{corbin2023assessing}, for example, higher $L^*$ leads to lighter skin color.} 
    \label{fig:Contributions of L and b to ITA value} 
\end{figure}

After the transformation to CIELAB space, Individual Typology Angle (ITA) of each image is computed based on formula~\ref{eq:ita}, which is a standard tool for Fitzpatrick Skin Type (FST) estimation~\cite{guo2023calibrating}, which is a scientific classification system based on the skin's reaction to ultraviolet (UV) radiation exposure.
\begin{equation}
    \mathit{ITA} = \arctan\left(\frac{L^* - 50}{b^*}\right) \times \frac{180}{\pi}
    \label{eq:ita}
\end{equation}
ITA value fully depends on $L^*$ and $b^*$. The contributions of them to skin color is shown in Figure~\ref{fig:Contributions of L and b to ITA value}. As shown in the figure, higher $L^*$ leads to lighter skin color.

\subsubsection{Select a target FST category}
The skin tone can be transformed in a random or targeted manner. For skin lesion detection, the sample number for darker skin tones is much lower than that of lighter skin tones, so selecting the former as target tone can be an effective way to reduce the bias. The mapping of FST category to skin tone is shown in table~\ref{tab:fst_ita}.
\begin{table}[htbp]
\centering
\caption{Fitzpatrick Skin Type (FST) Classification by ITA Values}
\label{tab:fst_ita}
\begin{tabular}{lll}
\toprule
\textbf{FST Type} & \textbf{ITA Range (degrees)} & \textbf{Description} \\
\midrule
I      & ITA > 55           & Very light skin \\
II     & 41 < ITA $\leq$ 55 & Light skin \\
III    & 28 < ITA $\leq$ 41 & Intermediate light skin \\
IV     & 10 < ITA $\leq$ 28 & Intermediate skin \\
V      & -30 < ITA $\leq$ 10 & Dark skin \\
VI     & ITA $\leq$ -30     & Very dark skin \\
\bottomrule
\end{tabular}
\end{table}
According to the table, darker skin type is in higher FST category. Therefore, selecting FST IV-VI can transform more dark skin to achieve the aim. In this method, FST IV is chosen as the target value.

\subsubsection{Adjust L and b to modify the skin tone}
In order to simulate the natural variability of the same FST in a real population and improve generalization capability, the target ITA value of each image will be randomly selected within the range of target FST. For a target ITA, there are infinite pairs of $L$ and $b$, leading to infinite transforming directions and local optimum. Therefore, the target $L$ and $b$ will be computed in two directions. First, $b$ will be fixed as $b_0$ and the only optimal $L_1$ can be calculated by formula 2. 
\begin{equation}
L_1 = b_0 \cdot \tan\left( \frac{\mathrm{ITA}_{\text{target}} \cdot \pi}{180} \right) + 50
\end{equation}
Second, $L$ will be fixed as $L_0$ and the only optimal $b_1$ can be calculated by formula 3. 
\begin{equation}
b_1 = \frac{L_0 - 50}{\tan\left( \frac{\mathrm{ITA}_{\text{target}} \cdot \pi}{180} \right)}
\end{equation}
Then there will be two solutions ($L_1$, $b_0$), ($L_0$, $b_1$), and the solution which is closer to original image will be selected as initial points. The final $L$ and $b$ will be optimized by gradient descent using the loss function to target ITA and a regularization term to keep the distance between transformed image to original image, which is shown in fomula 4.
\begin{equation}
\mathcal{L}(L, b) =
\underbrace{
\left[ \arctan\left( \frac{L - 50}{b} \right) \cdot \frac{180}{\pi}
- \mathrm{ITA}_{\text{target}} \right]^2}_{\text{ITA mismatch loss}}
+ \lambda_L (L - L_{\text{orig}})^2 + \lambda_b (b - b_{\text{orig}})^2
\end{equation}
After calculating the loss, $L$ and $b$ will be updated according to the gradient, which are shown in fomula 5 and fomula 6.
\begin{align}
L &\leftarrow L - \eta \cdot \frac{\partial \mathcal{L}}{\partial L} \\
b &\leftarrow b - \eta \cdot \frac{\partial \mathcal{L}}{\partial b}
\end{align}
\noindent where:
\begin{itemize}
    \item $\eta$ is scaling factor
\end{itemize}
The update will only be done once for each image in order to save time.

\subsubsection{Gaussian Filtering}
After applying color transformations on non-lesion skin area, the edge between the skin and the lesion may exhibit color breaks, making it appear unnatural. For example, the lesion edge may suddenly become brighter or yellower; and the transition area may show a noticeable boundary line. To solve this problem, the lesion mask is used to extract the lesion edge and a Gaussian filter is used for edge transition~\cite{basu2002gaussian}.
The 2D Gaussian kernel used for filtering is defined in fomula 7.
\begin{equation}
G(x, y) = \frac{1}{2\pi\sigma^2} \cdot \exp\left( -\frac{x^2 + y^2}{2\sigma^2} \right)
\end{equation}
\noindent where:
\begin{itemize}
    \item  $(x, y)$ is the spatial offset from the kernel center
    \item  $\sigma$ controls the degree of smoothing
\end{itemize}

The blurred image $I_{\text{blurred}}$ is obtained by convolving the original image $I$ with the Gaussian kernel based on fomula 8.
\begin{equation}
I_{\text{blurred}}(i, j) = \sum_{u=-k}^{k} \sum_{v=-k}^{k} G(u, v) \cdot I(i+u, j+v)
\end{equation}
\noindent where:
\begin{itemize}
    \item $k$ is the kernel radius, typically chosen so that kernel size is $(2k+1) \times (2k+1)$
\end{itemize}

Let $M_{\text{edge}}(i,j)$ be a binary mask indicating the boundary region between skin and lesion ($1$ on boundary, $0$ elsewhere). The blurred and original values are blended according to formula 9.
\begin{equation}
I_{\text{final}}(i, j) =
\begin{cases}
I_{\text{blurred}}(i,j), & \text{if } M_{\text{edge}}(i,j) = 1 \\
I(i,j), & \text{otherwise}
\end{cases}
\end{equation}
Or in vectorized form in formula 10
\begin{equation}
I_{\text{final}} = M_{\text{edge}} \cdot I_{\text{blurred}} + (1 - M_{\text{edge}}) \cdot I
\end{equation}

This operation is applied to $L$ and $b$ channels in Lab space after skin tone transformation which is shown in formula 11 and 12.
\begin{align}
I'_L &= M_{\text{edge}} \cdot (G * I_L) + (1 - M_{\text{edge}}) \cdot I_L \\
I'_b &= M_{\text{edge}} \cdot (G * I_b) + (1 - M_{\text{edge}}) \cdot I_b
\end{align}
\noindent where:
\begin{itemize}
    \item  $G * I$ denotes the convolution of the image $I$ with Gaussian kernel $G$.
\end{itemize}

Compared with original image, the  edge between lesion and skin area in the filtered image is more smooth and looks more nature, which can improve the training effects.

\subsubsection{Transformation Results}
\begin{figure}[htbp]
    \centering
    \begin{subfigure}[b]{0.45\textwidth}
        \includegraphics[width=5cm, height=5cm]{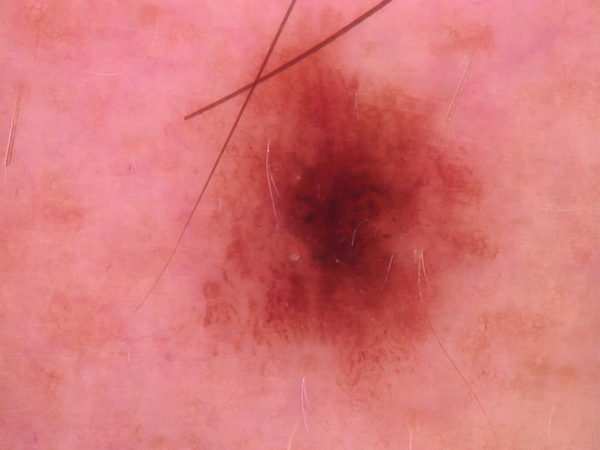} 
        \caption{Original Image}
        \label{fig:Original Image}
    \end{subfigure}
    \hfill 
    \begin{subfigure}[b]{0.45\textwidth}
        \includegraphics[width=5cm, height=5cm]{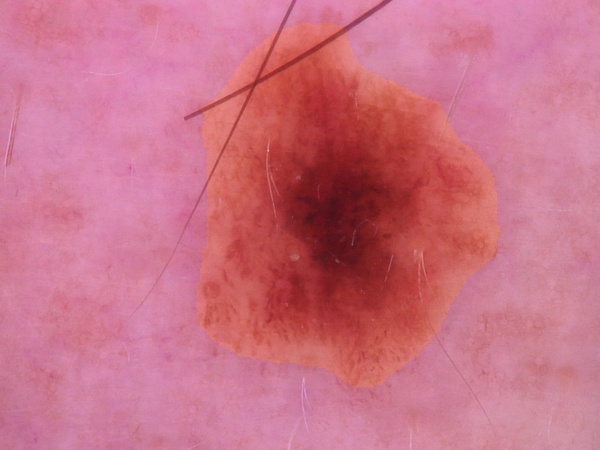} 
        \caption{Transformed Image}
        \label{fig:Transformed Image}
    \end{subfigure}
    \caption{The transformed image has different skin tone while the lesion area remains unchanged}
    \label{fig:Comparision}
\end{figure}

The transformed skin area will be combined with the original lesion area to form the final transformed image. A comparision between original image and transformed image is shown in Figure~\ref{fig:Comparision}. 

The figure shows that after skin tone normalization, the surrounding skin area transforms while the lesion area remains the same. 
\begin{table}[ht]
\centering
\caption{Fitzpatrick Skin Type (FST) distribution before and after skin tone transformation.}
\begin{tabular}{ccc}
\toprule
\textbf{FST Type} & \textbf{Transformed Count} & \textbf{Original Count} \\
\midrule
1 & 4   & 1    \\
2 & 4076 & 6619 \\
3 & 5831 & 3388 \\
4 & 104  & 7    \\
\bottomrule
\end{tabular}
\label{tab:fst_distribution}
\end{table}
The skin tone distribution after transformation is shown in table 2. The result shows that the number of FST2 decreases from 6619 to 4076, and that of FST3 increases from 3388 to 5831, making the skin tone distribution more balanced.

\subsubsection{Adaptive Blend Training}
After the skin tone normalization on each image, there will be two datasets, the originl and the transformed. To make full use of the augmented information, the original and transformed datasets are combined judiciously for model training. Adaptive blend training is used to further improve fairness on minor classes. The first step is initial probability calculation. Let $D_{\text{orig}}$ be the original dataset with FST distribution $P_{\text{orig}}(i)$ for $i \in \{1,...,6\}$. The probability of of replacing an original image with FST=$i$'s synthetic counterpart is defined in formula 13.
\begin{equation}
P_{\text{synth}}^{(i)} = \max\left(0, \frac{P_{\text{target}} - P_{\text{orig}}(i)}{P_{\text{target}}}\right)
\end{equation}
\noindent where:
\begin{itemize}
    \item  $P_{\text{target}} = 1/6$ represents the desired uniform distribution
\end{itemize}

During each training epoch, for every image $(x, y)$ in a mini-batch, draw a Bernoulli sample with probability $P_{\text{synth}}^{(y)}$. If the sample is positive, replace the original image $x$ with its corresponding synthetic version $x_{\text{synth}}$. Each synthetic image is generated in a one-to-one correspondence with the original, ensuring label consistency and visual alignment. Every $K$ epochs, the model's AUC is evaluated separately on each FST group. If the AUC for any group $i$ falls below a performance threshold $\tau$, its replacement probability is incrementally increased based on formula 14.
\begin{equation}
P_{\text{synth}}^{(i)} \gets \min(P_{\text{synth}}^{(i)} + \Delta, 0.9)
\end{equation}
This adaptive adjustment encourages the model to focus more on skin types with lower predictive performance, improving fairness and representation across all demographic groups. The pseudo-code is described in Algorithm~\ref{alg:adaptive}.
\begin{algorithm}[t]
\caption{Adaptive FST-Aware Data Blending}
\label{alg:adaptive}
\begin{algorithmic}[1]
\State \textbf{Input:} Original images $X_{\text{orig}}$, synthetic images $X_{\text{synth}}$, FST labels $Y_{\text{fst}}$
\State Initialize $P_{\text{synth}}^{(i)}$ using Eq. (13)
\For{epoch = 1 to $N$}
    \For{batch in DataLoader}
        \For{each $(x,y)$ in batch}
            \If{$\text{random}() < P_{\text{synth}}^{(y)}$}
                \State Replace $x$ with $x_{\text{synth}}$
            \EndIf
        \EndFor
        \State Train model on current batch
    \EndFor
    \If{epoch $\%$ $K$ == 0}
        \State Evaluate $AUC^{(i)}$ on validation set
        \For{each FST $i$}
            \If{$AUC^{(i)} < \tau$}
                \State $P_{\text{synth}}^{(i)} \gets \min(P_{\text{synth}}^{(i)} + \Delta, 0.9)$
            \EndIf
        \EndFor
    \EndIf
\EndFor
\end{algorithmic}
\end{algorithm}
The Baseline Parameters are defined as follows:
\begin{itemize}
\item Initial blending ratio: Computed from Eq. (13)
\item Adjustment threshold ($\tau$): 0.7 (for medical reliability)
\item Step size ($\Delta$): 0.05 (empirically determined)
\end{itemize}
Compared to fixed ratios, adaptive blend training can automatically adjust the FST balance and improve the fairness on minor classes.

\subsection{Joint Channel Pruning}
Skin tone normalization operates at the data processing stage, enhancing fairness by increasing sample diversity. Channel pruning helps to ensure fairness at the model design level. In convolutional neural networks (CNNs), different channels are sensitive to different features~\cite{el-khatib2020deep}. For example, some channel focus more on lesion area, while others pay more attention on skin tone. Therefore, cutting the channels which are more sensitive to skin tone can make the model focus more on lesion area, that is the intuition or mootivation for the channel pruning. 

\subsubsection{Deep Network}
The baseline model used for channel pruning is the standard Resnet50~\cite{he2016deep}, which is pretrained on ImageNet. There are two main advantages of using Resnet50 in skin lesion classification. The first is its deep architecture with high representational capacity. ResNet-50 is a 50-layer deep residual network capable of effectively capturing intricate features in dermatological images, such as complex textures, morphological patterns, and color variations. Its depth makes it particularly suitable for datasets like ISIC2019, which comprises multiple lesion categories with significant inter-class variability. The other advantage is its residual connections for gradient vanishing mitigation. This property is crucial for ISIC2019, where robust feature extraction is needed across diverse lesion types ranging from melanomas to benign nevi. 

\subsubsection{Training}
The baseline model is trained on ISIC2019 training dataset before pruning. The loss function used for multi-classification problem is cross-entropy loss. This function is widely used in deep learning, particularly for classification tasks~\cite{pundhir2023ethical}. It measures the difference between the predicted probability distribution and the true distribution. The formula is shown in equation 15.
\begin{equation}
\mathcal{L} = -\frac{1}{N}\sum_{i=1}^N \sum_{c=1}^C y_{i,c} \log(p_{i,c})
\end{equation}

where:
\begin{itemize}
    \item $y_{i,c}$ is the true label (one-hot encoded) for the $i$-th sample in class $c$
    \item $p_{i,c}$ is the model's predicted probability for the $i$-th sample belonging to class $c$
\end{itemize}
Ten fold cross-validation is used to avoid overfitting. The training dataset is randomly split to ten equal-sized folds. Each fold is used as validation set in turn, and the remaining nice folds are used as training set. The model is trained and validated for ten iterations. 

\subsubsection{Channel Pruning for single sensitive attribute}
\begin{figure}[htbp]
    \centering
    \includegraphics[width=0.8\textwidth]{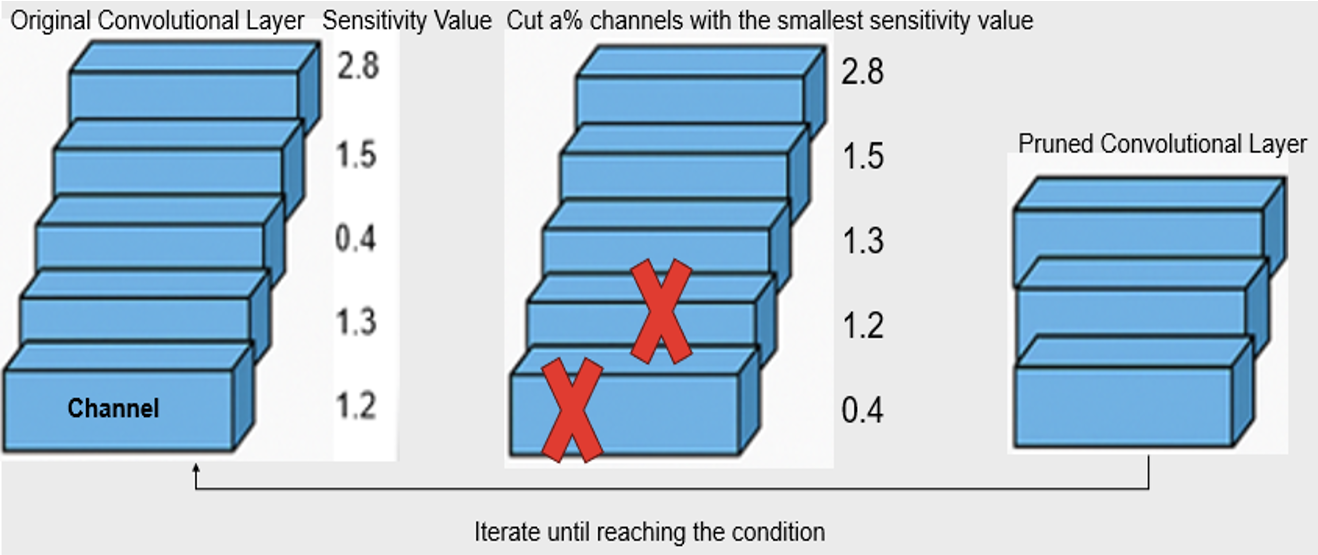} 
    \caption{Channel Pruning} 
    \label{fig:Channel Pruning} %
\end{figure}

In a CNN architecture, different layers are used to extract different features. The shallow layers usually extract low level features such as edge and texture, which are more general and not belong to specific populations. The deeper layers will extract high level features such as lesion shape and skin background, which are more sensitive to group difference~\cite{mustafa2025deep}. Therefore, the pruning will be done on the last convolutional layer of Resnet50. To prune the channels which are the most sensitive to skin tone, skin tone will be treated as the sensitive attribute $s$, and any sample $i$ will be assigned a sensitive attribute $s_i \in \{0,1\}$ (e.g., 0 represents light skin and 1 represents dark skin). For any sample $i$, any channel $k$ will extract the feature $f_i^k$. After feeding a batch of samples (suppose $b$ samples), we define the feature vector for channel $k$ as:
\[
\mathbf{F^k} = [f_1^k, f_2^2, \dots, f_b]^\top \in \mathbb{R}^{b \times 1}
\]
The sensitivity of the channel can be calculated by Soft Neatest Neighbor Loss (SNNL)~\cite{kong2024achieving}, which was originally proposed to quantify the \textit{intra-class compactness} and \textit{inter-class separation} in feature space. In pruning applications, it serves as a metric to evaluate whether a specific feature channel exhibits discriminative power for sensitive attributes. The equation of SNNL is shown in formula 16.
\begin{equation}
L_{SNNL} = \frac{1}{b} \sum_{i=1}^{b} \left[ -\log \left( 
\frac{
\sum\limits_{\substack{j\neq i}} e^{-\|\bm{f}_i - \bm{f}_j\|^2 / T} \cdot \mathbb{1}[s_i = s_j]
}{
\sum\limits_{\substack{j\neq i}} e^{-\|\bm{f}_i - \bm{f}_j\|^2 / T}
} \right) \right]
\label{eq:snnl}
\end{equation}

\noindent where:
\begin{itemize}
    \item $\bm{f}_i$ is the average feature vector of the $i$-th sample
    \item $s_i$ is the skin tone label (FST type) of the $i$-th sample
    \item $T$ is the temperature parameter controlling sensitivity to distance
    \item $b$ is the batch size (number of samples)
    \item $\mathbb{1}[\cdot]$ is the indicator function (1 when condition is true, 0 otherwise)
    \item $\|\cdot\|$ denotes the $L_2$ norm (Euclidean distance)
\end{itemize}
High SNNL score indicates high intra-class distance, which means the channel is insensitive to the specific attritube. Low SNNL score shows that the channel preserves discriminative power for senfitive attribute, which should be pruned. The pruning process is visualized in Figure~\ref{fig:Channel Pruning}.
It can be seen from the figure that the number of channels is reduced after one iteration of pruning.

\subsubsection{Fine tuning}
After channel pruning, the model parameters have been reduced and model structure has been changed. Some channels which were previously used for feature extraction or classification in training stage have been dropped. Therefore, the model needs to be fine tuned after each iteration of pruning in order to adapt to new structure and recover model performance. Due to the pruning of channels, the output dimension of the pruned layer has been changed, so the first step of fine tuning is to adjust the input dimension of the next layer to make the whole model structure consistent. The process of channel pruning together with fine tuning is presented in Algorithm~\ref{alg:pruning}. Though the training time will increase due to fine tuning after each iteration, the inference time can be lower because of the reduction of model parameters. This reduction in inference time is an advantage for deployment in low resource clinical settings since the training can be performed offline a-priori in a high compute environment.
\begin{algorithm}
\caption{Iterative SNNL-Guided Pruning with Fine-tuning}
\label{alg:pruning}
\begin{algorithmic}[1]
\Statex \textbf{Input:} 
\State \quad Pretrained model $M_0$ with parameters $\Theta_0$
\State \quad Training data $D_{\text{train}} = \{(x_i, y_i, s_i)\}_{i=1}^N$ (images, labels, sensitive attributes)
\State \quad Pruning ratio $\prc$ per iteration (default: 2\%)
\State \quad Max iterations $n_{\text{max}}$ (default: 3)
\State \quad Accuracy threshold $\tacc$ (e.g., 3\% drop)
\State \quad Fairness threshold $\tfair$ (e.g., 0.5\% improvement)
\Statex \textbf{Output:} Pruned model $M^*$ with improved fairness

\For{$k = 1$ \textbf{to} $n_{\text{max}}$}
    \State \textbf{Channel Evaluation:}
    \For{\textbf{each} channel $c$ \textbf{in} target layers}
        \State Compute SNNL-Fair score:
        \[
        \gamma_c = \frac{1}{|B|}\sum_{i\in B} -\log\left(
        \frac{\sum_{j\neq i} \exp(-\frac{\|f_i^c - f_j^c\|^2}{T})\cdot \mathbb{I}(s_i=s_j)}
        {\sum_{j\neq i} \exp(-\frac{\|f_i^c - f_j^c\|^2}{T})} \right)
        \]
    \EndFor
    
    \State \textbf{Pruning:}
    \State Sort channels by $\gamma_c$ in ascending order
    \State Select top $\prc\%$ channels with lowest $\gamma_c$ as $C_{\text{prune}}$
    \State Update model: $M_k \leftarrow \text{PruneChannels}(M_{k-1}, C_{\text{prune}})$
    
    \State \textbf{Fine-tuning:}
    \While{not converged}
        \For{\textbf{each} batch $(X_b, Y_b, S_b)$ \textbf{in} $D_{\text{train}}$}
            \State Update parameters: $\Theta_k \leftarrow \Theta_k - \eta \nabla_\Theta \mathcal{L}_{\text{CE}}(M_k(X_b), Y_b)$
        \EndFor
    \EndWhile
    
    \State \textbf{Termination Check:}
    \State Evaluate accuracy $A_k$ and fairness metric $F_k$ on validation set
    \If{$A_0 - A_k > \tacc$ \textbf{or} $F_k - F_{k-1} < \tfair$}
        \State \Return $M_{k-1}$ \Comment{Revert to previous model}
    \EndIf
\EndFor
\State \Return $M_k$
\end{algorithmic}
\end{algorithm}

\subsubsection{Channel Pruning for multiple sensitive attribute}
To improve model fairness across multiple sensitive attributes (e.g., skin tone, gender, age), the initial pruning method is extended to a meta learning based channel pruning framework. To consider multiple sensitive attributes, the total sensitive loss function becomes the weighted sum of SNNL loss on each sensitive attributes. However, it is difficult to decide correct weights at the beginning. Therefore, instead of assigning fixed weights to each attribute's fairness loss, the weights are treated as learnable parameters and they will be optimized based on the model fairness performance on a held-out meta set which is 10\% split from the training set.
Specifically, a vector of learnable fairness weights is defined as follows:
\[
\mathbf{w} = [w_1, w_2, \dots, w_K] \in \mathbb{R}^K, \quad w_i \geq 0
\]
where $K$ is the number of sensitive attributes (in this case, $K=3$). These weights are used to compute a weighted SNNL on the training set:
\[
\mathcal{L}_{\text{train}}(\theta; \mathbf{w}) = \sum_{i=1}^{K} w_i \cdot \text{SNNL}(f_\theta(x), a_i)
\]
Here, $f_\theta$ denotes the feature extractor with parameters $\theta$, and $a_i$ is the $i$-th sensitive attribute. Given a candidate set of weights $\mathbf{w}$, the updated model parameters after one step of pruning-aware training is obtained as follows:
\[
\theta' = \theta - \alpha \nabla_\theta \mathcal{L}_{\text{train}}(\theta; \mathbf{w})
\]

Then, the model is evaluated on the meta set using Group-wise Variance Fairness Loss. Given a sensitive attribute $a$ with $G$ discrete groups (e.g., Male/Female, Light/Dark skin), the group-wise average loss for group $g$ is defined as:
\[
\ell_g = \frac{1}{|\mathcal{D}_g|} \sum_{(x, y) \in \mathcal{D}_g} \mathcal{L}(f_{\theta'}(x), y), \quad g = 1, \dots, G
\]
\noindent where:
\begin{itemize}
    \item  $\mathcal{D}_g = \{(x, y) \in \mathcal{D}_{\text{meta}} \mid a(x) = g\}$ is the subset of meta data belonging to group $g$
    \item  $\mathcal{L}$ is the task loss (e.g., cross-entropy)
\end{itemize}
The group-wise variance fairness loss is then defined as:
\[
\mathcal{L}_{\text{meta}}^{\text{fair}} = \text{Var}(\ell_1, \ell_2, \dots, \ell_G) = \frac{1}{G} \sum_{g=1}^{G} (\ell_g - \bar{\ell})^2
\]
\noindent where:
\begin{itemize}
    \item  $\bar{\ell} = \frac{1}{G} \sum_{g=1}^{G} \ell_g$ is the mean group loss. 
\end{itemize}

\noindent
A smaller $\mathcal{L}_{\text{meta}}^{\text{fair}}$ indicates more equitable performance across groups, and this loss is differentiable, enabling backpropagation to upstream fairness weights.

\[
\mathcal{L}_{\text{meta}}^{\text{total}} = \sum_{k=1}^{K} \text{Var}(\ell_1^{(k)}, \dots, \ell_{G_k}^{(k)})
\]

Finally, $\mathbf{w}$ is optimized using gradients from the meta loss:
\[
\mathbf{w} \leftarrow \mathbf{w} - \eta \nabla_\mathbf{w} \mathcal{L}_{\text{meta}}(\theta')
\]
The full process of joint channel pruning is shown in Algorithm~\ref{alg:meta_pruning_var}.

\begin{algorithm}[H]
\caption{Meta-Learning-Based Fairness-Aware Channel Pruning (Group-wise Variance Loss)}
\label{alg:meta_pruning_var}
\begin{algorithmic}[1]
\Require Initial model parameters $\theta$, learning rates $\alpha$, $\eta$
\Require Training set $\mathcal{D}_{\text{train}}$, Meta set $\mathcal{D}_{\text{meta}}$
\Require Sensitive attributes $a_1, \dots, a_K$ with group labels
\State Initialize fairness weights $\mathbf{w} \gets [1.0, \dots, 1.0]$
\For{$t = 1$ to $T$} \Comment{Meta-learning iterations}
    \State Sample mini-batch $(x, a)$ from $\mathcal{D}_{\text{train}}$
    \State Compute weighted SNN loss:
    \[
    \mathcal{L}_{\text{train}} = \sum_{i=1}^{K} w_i \cdot \text{SNNL}(f_\theta(x), a_i)
    \]
    \State Simulate parameter update:
    \[
    \theta' \gets \theta - \alpha \nabla_\theta \mathcal{L}_{\text{train}}
    \]
    \State Sample meta mini-batch $(x_{\text{meta}}, y_{\text{meta}}, a_{\text{meta}})$ from $\mathcal{D}_{\text{meta}}$
    \State Compute meta fairness loss as group-wise variance:
    \For{$i = 1$ to $K$}
        \State Partition $(x_{\text{meta}}, y_{\text{meta}})$ into groups $\{\mathcal{D}^{(i)}_g\}_{g=1}^{G_i}$ based on $a_i$
        \For{$g = 1$ to $G_i$}
            \State Compute group loss: $\ell^{(i)}_g = \frac{1}{|\mathcal{D}^{(i)}_g|} \sum_{(x, y) \in \mathcal{D}^{(i)}_g} \mathcal{L}(f_{\theta'}(x), y)$
        \EndFor
        \State Compute variance: $\mathcal{L}^{(i)}_{\text{meta}} = \text{Var}(\ell^{(i)}_1, \dots, \ell^{(i)}_{G_i})$
    \EndFor
    \State Aggregate: $\mathcal{L}_{\text{meta}} = \sum_{i=1}^{K} \mathcal{L}^{(i)}_{\text{meta}}$
    \State Update fairness weights:
    \[
    \mathbf{w} \gets \mathbf{w} - \eta \nabla_\mathbf{w} \mathcal{L}_{\text{meta}}
    \]
    \State Project $\mathbf{w}$ onto positive space (optional): $w_i \gets \max(0, w_i)$
\EndFor
\State Use final $\mathbf{w}$ to compute channel importance via weighted SNNL
\State Prune channels with lowest fairness scores
\end{algorithmic}
\end{algorithm}
With these two inserted loops, the optimal weight of each sensitive attribute can be found and the fairness can be improved on multi-dimensions. This method can be expanded to include additional sensitive attributes beyond skin tone, age, and gender.

\subsection{Interpretability}
The model interpretability can be improved by class activation mapping and visualization of feature map.
\subsubsection{CAM}
Class Activation Mapping (CAM) and Gradient-weighted Class Activation Mapping (Grad-CAM) are visualization techniques that highlight the important regions in an image towards predicting a particular class in convolutional neural networks~\cite{zhou2016cam}. CAM requires a specific CNN architecture:
\begin{itemize}
    \item Convolutional layers + Global Average Pooling (GAP) layer
    \item Fully connected layer with softmax activation at the end
\end{itemize}
For a given class $c$, the CAM $L_{CAM}^c$ is computed in formula 17.
\begin{equation}
L_{CAM}^c(x,y) = \sum_k w_k^c \cdot f_k(x,y)
\end{equation}
Where:
\begin{itemize}
    \item $f_k(x,y)$ is the activation of unit $k$ in the last convolutional layer at spatial location $(x,y)$
    \item $w_k^c$ is the weight connecting the $k$-th feature map to the output class $c$, which come from the fully connected layer after global average pooling
\end{itemize}
The limitation of CAM is that it requires specific architecture and it needs model retraining before visualization. This method cannot be applied to arbitrary CNN architectures. Compared with CAM, Grad-CAM can generalize to any CNN architecture. It uses gradient information flowing into the last convolutional layer, which does not require architectural changes and it works with any CNN-based model~\cite{selvaraju2017grad}. For a given class $c$, the Grad-CAM $L_{Grad-CAM}^c$ is computed as formula 18.
\begin{equation}
L_{Grad-CAM}^c = ReLU\left(\sum_k \alpha_k^c \cdot A^k\right)
\end{equation}
Where:
\begin{itemize}
    \item $A^k$ is the activation of the $k$-th feature map in the last convolutional layer
    \item $\alpha_k^c$ are the neuron importance weights computed as:
\end{itemize}

\begin{equation}
\alpha_k^c = \frac{1}{Z}\sum_i\sum_j \frac{\partial y^c}{\partial A_{ij}^k}
\end{equation}

\begin{itemize}
    \item $Z$ is the number of pixels in the feature map
    \item $\frac{\partial y^c}{\partial A_{ij}^k}$ is the gradient of the score for class $c$ with respect to the activation at location $(i,j)$ in feature map $k$
\end{itemize}
For sharper visualizations, Grad-CAM can be combined with guided backpropagation as shown in formula 20.

\begin{equation}
L_{Guided\ Grad-CAM}^c = L_{Grad-CAM}^c \odot L_{Guided\ Backprop}^c
\end{equation}
\noindent where:
\begin{itemize}
    \item  $\odot$ denotes element-wise multiplication. 
\end{itemize}

\subsubsection{Visualization of pruned channel}
The other way to improve interpretability is to visualize the process of fairness improvement in channel pruning. For each image, the output of each channel called feature map shows the activation distribution on that image. In other words, the feature map shows the attention of the channel on a given image. The dropped channels in each iteration will focus on more unrelated areas such as skin background on a batch of input images, thus the corresponding feature maps will have higher activation values in background area. Therefore, the feature maps of pruned channels can be saved and visualized to explain why pruning these channels is correct. The process of visualization is shown in Algorithm~\ref{alg:Visualize Sensitive Channels}.
\begin{algorithm}
\caption{Visualize Sensitive Channels: Overlay Heatmaps of Sensitive Channels on Input Images}
\label{alg:Visualize Sensitive Channels}
\begin{algorithmic}[1]
\Require Trained model $M$, DataLoader $\mathcal{D}$, sensitive channel indices $\mathcal{C}$, device, output path $P$
\Ensure Saved overlay images of each sensitive channel per input image

\ForAll{batch $(X, Y, S, I)$ in $\mathcal{D}$}
    \State Move $X$ and $M$ to target device
    \State Compute feature maps $F = \text{extract\_feature\_resnet}(X)$ \Comment{$F \in \mathbb{R}^{B \times C \times H \times W}$}
    
    \For{$i = 1$ to batch size $B$}
        \State $x_i \gets X[i]$ \Comment{Input image}
        \State $f_i \gets F[i]$ \Comment{Feature map of image $i$}
        
        \ForAll{channel index $c$ in $\mathcal{C}$}
            \State $h \gets f_i[c]$ \Comment{Select feature map for channel $c$}
            \State Normalize $h$ to $[0,1]$ and resize to $224 \times 224$
            \State Generate heatmap $H_c$ using a color map
            \State Convert $x_i$ to RGB numpy image $I_i$
            \State Overlay $H_c$ on $I_i$ to get visualization $\hat{I}_c$
            \State Save $\hat{I}_c$ as \texttt{img\_i\_ch\_c.png} in folder $P$
        \EndFor
    \EndFor
\EndFor
\end{algorithmic}
\end{algorithm}

\section{Results}
All methods are tested on ISIC2019 dataset and all pruning methods are multiple sensitive attribute pruning. To test the fairness on different sensitive attributes, each image is assigned with three sensitive attribute labels. Each image is assigned with a Fitzpatrick Type among 2-5 according to its ITA value and an age group.
\subsection{Accuracy on Different Groups}
\begin{figure}[htbp]
    \centering
    \begin{subfigure}[t]{0.45\textwidth}
        \centering
        \includegraphics[width=\linewidth, height=5cm, keepaspectratio]{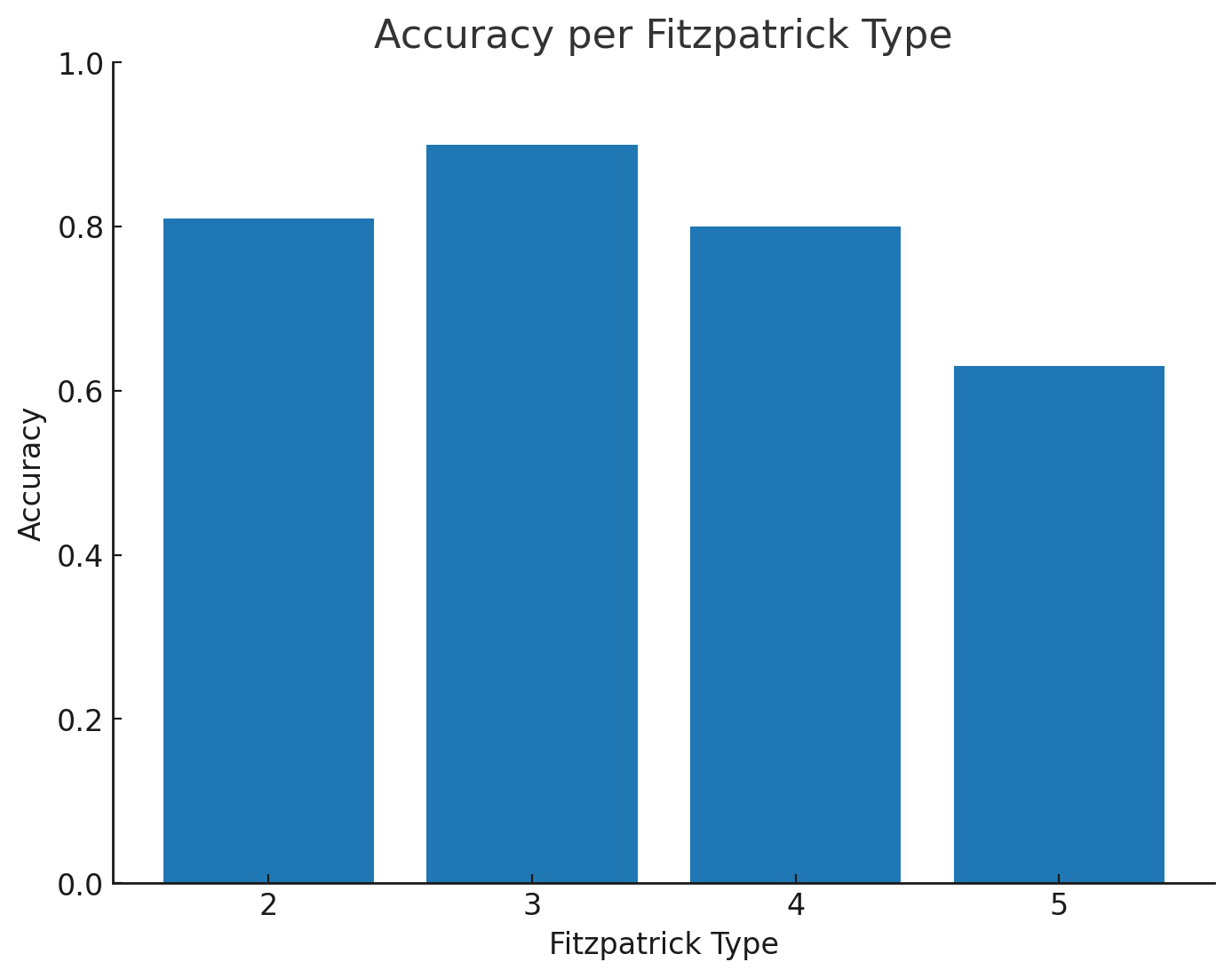}
        \caption{Accuracy per Fitzpatrick}
        \label{fig:Accuracy per Fitzpatrick}
    \end{subfigure}
    \hfill
    \begin{subfigure}[t]{0.45\textwidth}
        \centering
        \includegraphics[width=\linewidth, height=5cm, keepaspectratio]{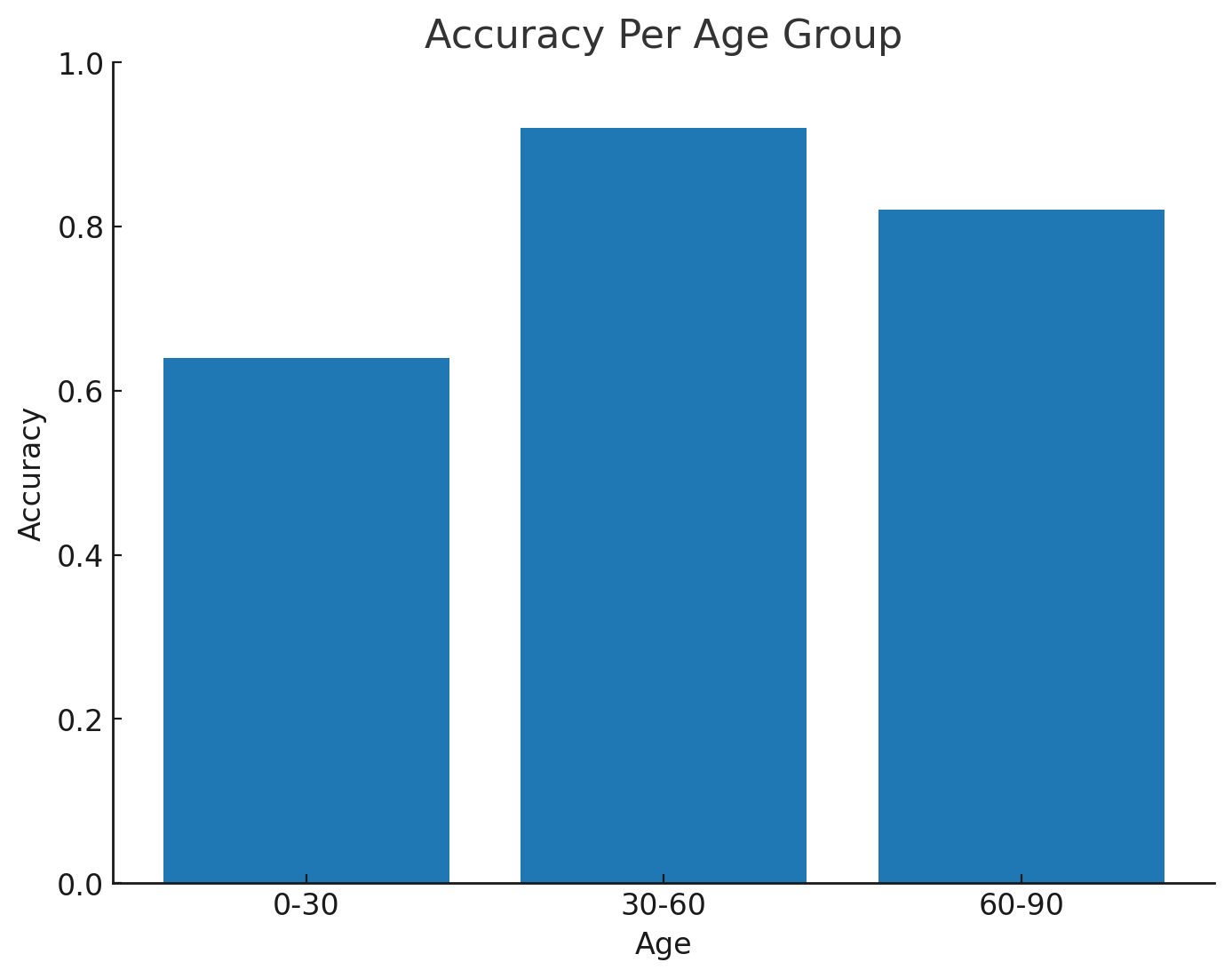}
        \caption{Accuracy per Age Group}
        \label{fig:Accuracy per Age Group}
    \end{subfigure}
    
    \vspace{0.5cm}  
    
    \begin{subfigure}[t]{0.45\textwidth}
        \centering
        \includegraphics[width=\linewidth, height=5cm, keepaspectratio]{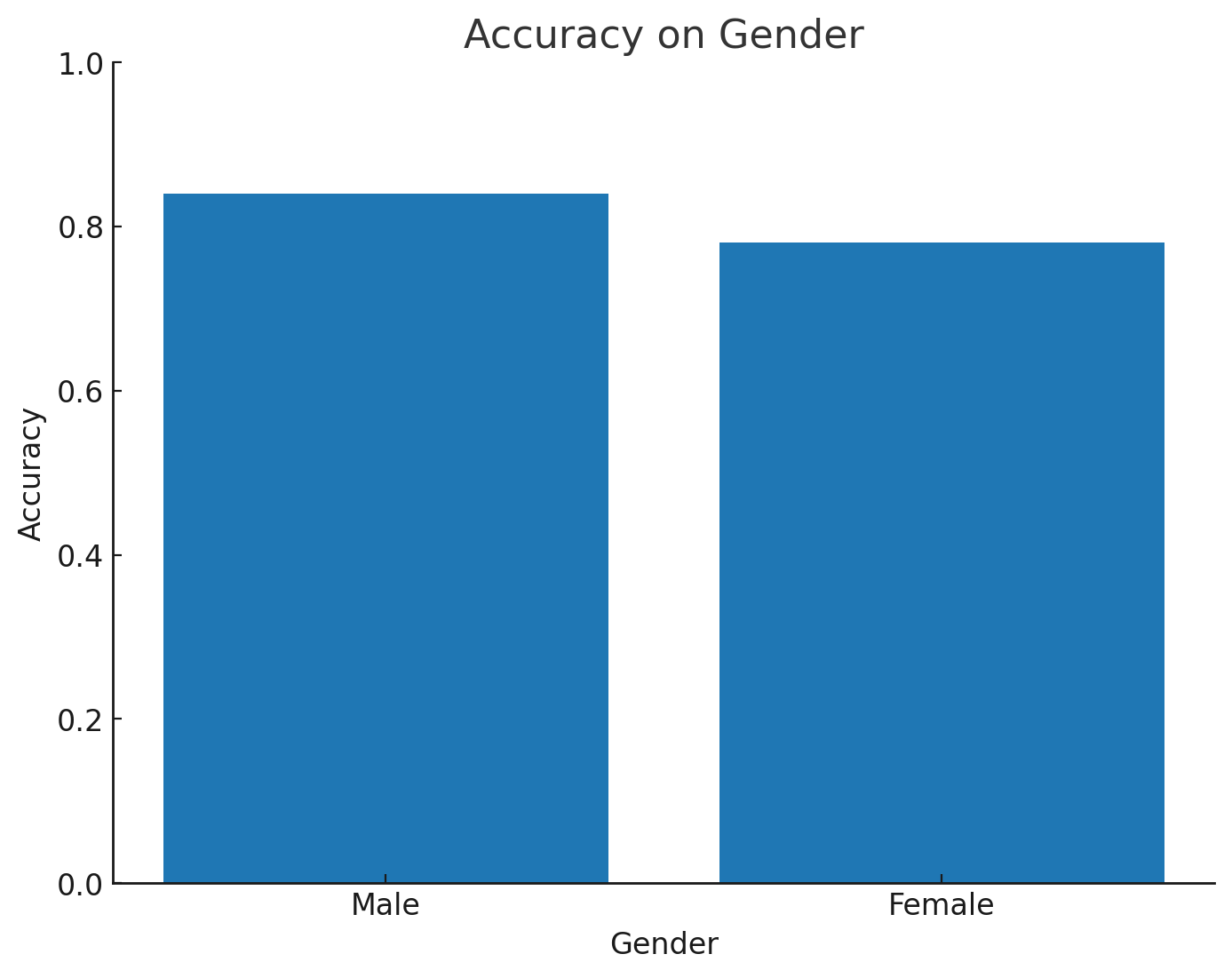}
        \caption{Accuracy on Gender}
        \label{fig:Accuracy on Gender}
    \end{subfigure}
    \caption{Accuracy on Different Groups}
    \label{fig:Accuracy on Different Groups}
\end{figure}

The accuracy on different groups is shown in Figure~\ref{fig:Accuracy on Different Groups}. 
According to the figure, the overall accuracy of the proposed method is over 80\%. In terms of skin tone, patients in Fitzpatrick type 2-4 are well classified, but the accuracy on patients in Fitzpatrick type 5 is lower, which means the model still have some residual biases on darker skin tone, but the performance is much better than baseline model. According to the results on age group, the accuracy reaches 90\% on 30-60 age group, while the accuracy is relatively low on patients below 30 years old, perhaps due to less number of training samples available for that group. The classification performance is well balanced across genders. 

\subsection{Ablation study}
As ablation study of the proposed approach, four combinations including the baseline model, baseline model with skin tone normalization, baseline model with channel pruning, baseline model with skin tone normalization as well as channel pruning are evaluated and the results are presented in table~\ref{tab:Performance Comparison of Different Methods}. 

\begin{table}[htbp]
\centering
\caption{Performance Comparison of Different Methods}
\label{tab:Performance Comparison of Different Methods}
\begin{tabular}{lccccccc}
\toprule
Method & Accuracy & F1-score & \makecell{EOdds \\ of skin tone} & \makecell{EOdds \\ of age} & \makecell{EOdds \\ of gender} & DI & ABROCA \\
\midrule
Baseline & 0.7273 & 0.7238 & 0.7037 & 0.6825& 0.7543& 0.8333 & 0.5 \\
With Color Transformation & 0.7931 & 0.7890 & 0.4556 & 0.6749& 0.7520& 0.8667 & 0.2415 \\
With Channel Pruning & 0.7340 & 0.7338 & 0.4930 & 0.4736& 0.4825& 0.9091 & 0.2117 \\
With Both & 0.8069 & 0.7947 & 0.4142 & 0.4728& 0.4823& 0.9167 & 0.1973 \\
\bottomrule
\end{tabular}
\end{table}

\subsection{Fairness evaluation}
\subsubsection{Equalized Odds}
Equalized Odds (EOdds)~\cite{hardt2016equality} is a common fairness metric which requires that
the true positive rate (TPR) and false positive rate (FPR) are equal across all groups defined by a sensitive attribute such as skin tone, age, and gender. The EOdds difference is defined in formula 21.
\begin{equation}
\text{Eodds Difference} = \frac{1}{2} \left( |TPR_{a} - TPR_{b}| + |FPR_{a} - FPR_{b}| \right)
\end{equation}
Where:
\begin{itemize}
    \item $TPR_{a} = P(\hat{Y} = 1 | A = a, Y = 1)$ is the true positive rate for group $a$
    \item $FPR_{a} = P(\hat{Y} = 1 | A = a, Y = 0)$ is the false positive rate for group $a$
\end{itemize}
Smaller EOdds difference indicates better fairness across groups, and the ideal value shoule be 0.

\subsubsection{ABROCA}
The other metric used here for fairness evaluation is Area Between ROC Curves (ABROCA), which measures algorithmic fairness by comparing classifier performance across different demographic groups~\cite{borchers2025abroca}. Unlike single-threshold metrics, ABROCA evaluates disparity across all possible classification thresholds. It is defined in formula 26. Compared with EOdds, ABROCA can quantify the severity of unfairness.
\begin{equation}
\text{ABROCA} = \int_{0}^{1} |ROC_A(f) - ROC_B(f)| \, df
\end{equation}
where:
\begin{itemize}
    \item $ROC_A(f)$ is the true positive rate for group A at false positive rate $f$
    \item $ROC_B(f)$ is the true positive rate for group B at false positive rate $f$
\end{itemize}
Small ABROCA value indicates good fairness, and the ideal value should be close to 0.

\begin{figure}[htbp]
    \centering
    \begin{subfigure}[b]{0.45\textwidth}
        \includegraphics[width=5cm, height=5cm]{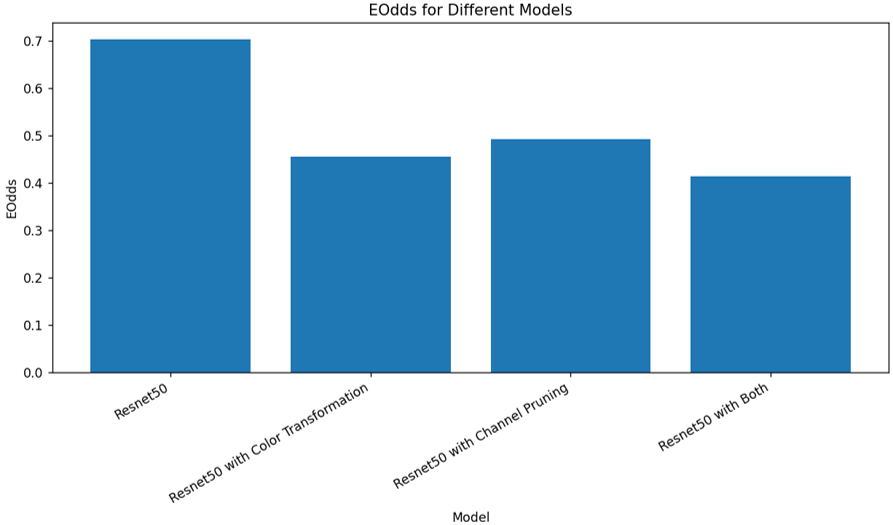} 
        \caption{EOdds}
        \label{fig:EOdds}
    \end{subfigure}
    \hfill 
    \begin{subfigure}[b]{0.45\textwidth}
        \includegraphics[width=5cm, height=5cm]{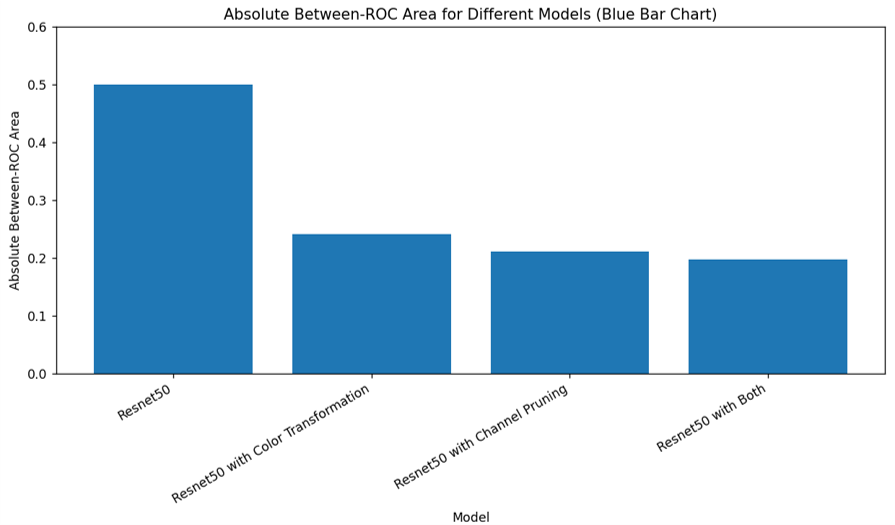} 
        \caption{ABROCA}
        \label{fig:ABROCA}
    \end{subfigure}
    \caption{Performance of Different Methods}
    \label{fig:Performance of Different Methods}
\end{figure}

The comparsion of EOdds on skin tone and ABROCA of different methods are visualized in Figure~\ref{fig:Performance of Different Methods}.

According to the results, skin tone normalization can significantly improve accuracy and fairness on skin tone, while it has few effects on the fairness of age and gender. Channel pruning has little influence on accuracy, while it can synthetically improve the fairness on skin tone, age, and gender. Thus the improvement on accuracy and fairness of the two proposed methods are complementary and the combination of them gives the best performance. 

\subsection{Qualitative results}
\begin{figure}[htbp]
    \centering
    \begin{subfigure}[b]{0.45\textwidth}
        \includegraphics[width=5cm, height=5cm]{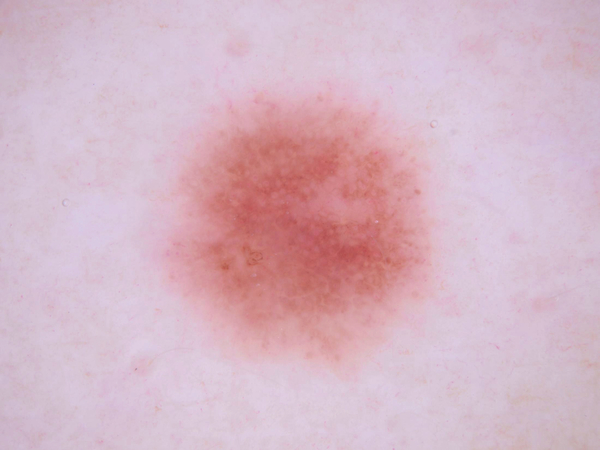} 
        \caption{input image}
        \label{fig:input image1}
    \end{subfigure}
    \hfill 
    \begin{subfigure}[b]{0.45\textwidth}
        \includegraphics[width=5cm, height=5cm]{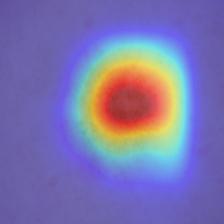} 
        \caption{heat map}
        \label{fig:heat map1}
    \end{subfigure}
    \caption{In the heat map of correct prediction, the lesion structure and it is well aligned with the original image}
    \label{fig:Heat map of correct prediction}
\end{figure}

\begin{figure}[htbp]
    \centering
    \begin{subfigure}[b]{0.45\textwidth}
        \includegraphics[width=5cm, height=5cm]{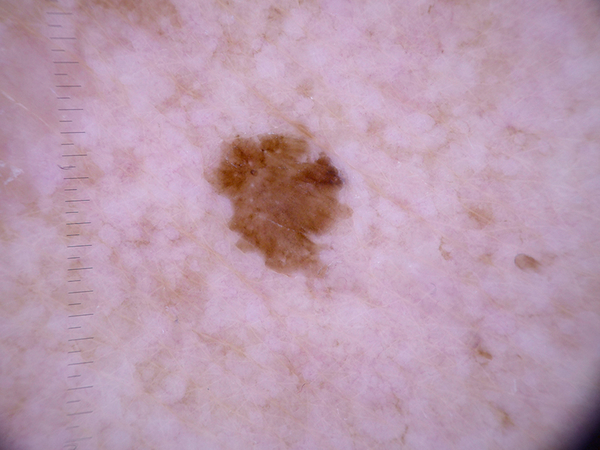} 
        \caption{input image}
        \label{fig:input image2}
    \end{subfigure}
    \hfill 
    \begin{subfigure}[b]{0.45\textwidth}
        \includegraphics[width=5cm, height=5cm]{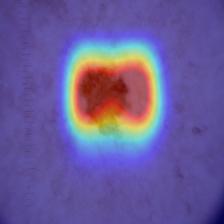} 
        \caption{heat map}
        \label{fig:heat map2}
    \end{subfigure}
    \caption{In the heat map of wrong prediction, there is a huge gap between the heatmap structure and the original image}
    \label{fig:Heat map of wrong prediction}
\end{figure}
Figure~\ref{fig:Heat map of correct prediction} shows a correct classified image and its corresponding heatmap generated by CAM. In the heatmap, more red-shifted area represents the higher probability the model considers that the area contributes to a specific disease. This heatmap clearly shows the lesion structure and it is well aligned with the original image. Figure~\ref{fig:Heat map of wrong prediction} shows the heatmap of a misclassified image. It can be seen that there is a huge gap between the heatmap structure and the original image with respect to the lesion shape, which means the model does not use correct localised information to make the decision.

\begin{figure}[htbp]
    \centering
    \begin{subfigure}[b]{0.18\textwidth}
        \includegraphics[width=\textwidth]{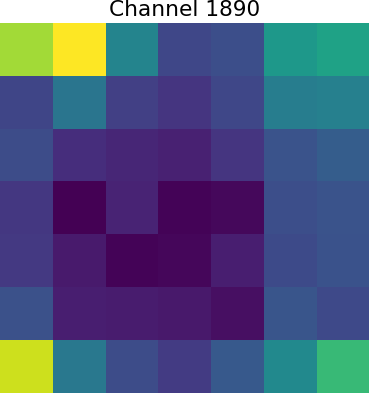}
        \caption{1}
    \end{subfigure}
    \hfill
    \begin{subfigure}[b]{0.18\textwidth}
        \includegraphics[width=\textwidth]{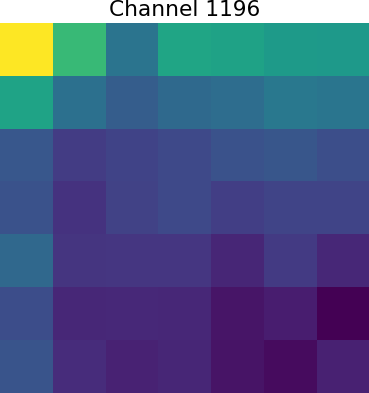}
        \caption{2}
    \end{subfigure}
    \hfill
    \begin{subfigure}[b]{0.18\textwidth}
        \includegraphics[width=\textwidth]{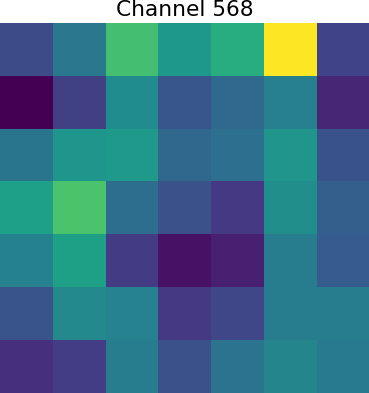}
        \caption{3}
    \end{subfigure}
    \hfill
    \begin{subfigure}[b]{0.18\textwidth}
        \includegraphics[width=\textwidth]{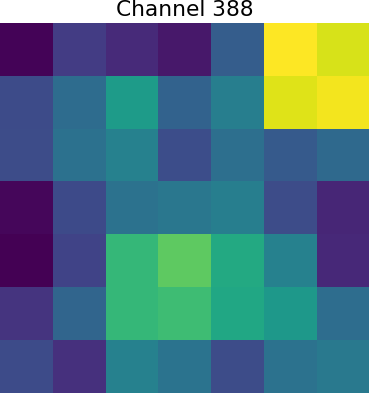}
        \caption{4}
    \end{subfigure}
    \hfill
    \begin{subfigure}[b]{0.18\textwidth}
        \includegraphics[width=\textwidth]{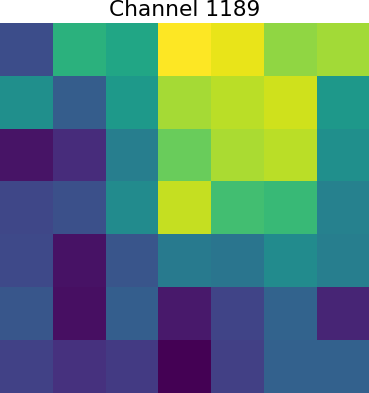}
        \caption{5}
    \end{subfigure}
    \caption{Feature maps of five example pruned channels}
    \label{fig:pruned_channels}
\end{figure}
Figure\ref{fig:pruned_channels} shows a visualization of feature maps of five examples of pruned channels successively. In each image, the lighter area represents a higher activation value, which means the channel is more focused on that area. For the first heatmap, the yellow area only appears at corners, which means that the channel only pays attention to the outside area instead of the lesion area in the center. This channel is only influenced by the skin background and will contribute to an unfair decision, so it should definitely be pruned. From the second to fourth images, the yellow area gradually converges from the periphery toward the center, which means that the channels become more focused on the lesion area after several iterations. In the final pruning iteration, as shown in the last image, the channel is more focused on lesion area than skin background, so the pruning should be stopped. In general, the improvement of performance can be explained with the visualization of the series of feature maps of pruned channels.

\subsection{Quantitative performance results against competing methods}
The proposed method is compared with five existing methods, which are all focused on fair detection of skin lesion and all methods are tested on ISIC2019 testing dataset under similar experimental conditions as far as possible. The comparative results are shown in table~\ref{tab:prior work}. 
\begin{table}[htbp]
\centering
\caption{Comparision}
\label{tab:prior work}
\begin{tabular}{lcccc}
\toprule
\textbf{Methods} & \textbf{Accuracy} & \textbf{F1-score} & \textbf{Eodd sex} & \textbf{Eodd skin} \\
\midrule
HSIC~\cite{quadrianto2019discovering}          & 0.731  & 0.7    & 0.02   & 0.166 \\
MFD~\cite{jung2021fair}           & 0.77   & 0.735  & 0.024  & 0.166 \\
Advconf~\cite{wu2023fairprune}       & 0.733  & 0.736  & 0.037  & 0.169 \\
DomainIndep~\cite{xu2025incorporating}   & 0.727  & 71.8   & 0.042  & 0.161 \\
FairAdaBN~\cite{xu2023fairadabn}     & 0.767  & 0.743  & 0.019  & 0.171 \\
Ours          & 0.75   & 0.732  & 0.02   & 0.148 \\
\bottomrule
\end{tabular}
\end{table}
It can be readily observed that though the accuracy of proposed method is at a similar level to our competitors, our fairness results across multiple sensitive attributes including skin tone and gender outperforms the others. This is because our joint pruning method can consider multiple attributes at the same time and while it may slightly reduce the performance on accuracy (still competitive though), the better fairness on skin tone makes it worthwhile as it increases the trustability of the approach to dermatologists for possible clinical adoption with higher assurance.

\section{Conclusion}
We demonstrate how fairness considerations can be directly integrated into optimal model design by iteratively pruning those channels that have less sensitivity to the task at hand with respect to several fairness sensitive attributes simultaneously. We evaluate the approach in the relevant application of skin lesion image classification where class imbalance is prevalent both with respect to lesion type as well as skin tone, as skin cancer is more prevalent among lighter skinned patients due to less melanin pigment concentration. The proposed methodology not only maintains a comparable accuracy with respect to competing state of the art methods, but it also outperforms the competitors with respect to fairness measures across patient subgroups with respect to gender, age, etc. Thus our method provides a layer of trustworthiness that might provide the clinical users to adopt it for screening use subject to future clinical validation studies. \\

One limitation of the current work is that there is limited public data available that supports this kind of thorough fairness studies in skin cancer. We expect to liaise with dermatologists at UCLH and University of Southampton in the future to validate the method on real life data. Also as part of future work, we would like to explore generative model (GAN diffusion, etc) based data augmentation for the skin tone normalization stage. For this work, we used a relatively simple but effective Fitzpatrick scale based transform technique. Channel Pruning can be extended to Multi-Objective Optimization problem such as Pareto optimization to find the optimal pruning scheme in order to balance the accuracy and fairness.

\section*{Author declarations and acknowledgments}
The authors declare no conflict of interest. TC is supported by a principal research fellowship from the UCL NIHR Biomedical Research Centre. He is also supported by the Turing-Roche Strategic Partnership between Roche Pharmaceuticals and the Alan Turing Institute, UK's national institute for AI.

% \newpage


\begin{thebibliography}{8}
\bibitem{Ali2022}
Karar Ali and Zaffar Ahmed Shaikh and Abdullah Ayub Khan and Asif Ali Laghari: Multiclass skin cancer classification using EfficientNets – a first step towards preventing skin cancer. Neuroscience Informatics, vol.2, p.100034 (2022)

\bibitem{Kalb2023}
Thorsten Kalb and Kaisar Kushibar: Revisiting Skin Tone Fairness in Dermatological Lesion Classification, arXiv, vol.2308.09640 (2023)

\bibitem{doshi2017towards}
Doshi-Velez, Finale and Kim, Been: Towards a rigorous science of interpretable machine learning, arXiv preprint arXiv:1702.08608 (2017)

\bibitem{guidotti2018survey}
Guidotti, Riccardo and Monreale, Anna and Ruggieri, Salvatore and Turini, Franco and Giannotti, Fosca and Pedreschi, Dino: A survey of methods for explaining black box models, ACM Computing Surveys (CSUR), vol.51, pp.1--42 (2018)

\bibitem{kamiran2009preprocessing}
Faisal Kamiran and Toon Calders: Data Preprocessing Techniques for Classification Without Discrimination, Proceedings of the 2009 IEEE International Conference on Computer, Control and Communication (IC4), pp.1--6, (2009)

\bibitem{agarwal2018reductions}
Alekh Agarwal and Alina Beygelzimer and Miroslav Dudík and John Langford and Hanna Wallach: A Reductions Approach to Fair Classification, Proceedings of the 35th International Conference on Machine Learning (ICML), vol.80, pp.60--69, (2018)

\bibitem{Yao2024}
Siqiong Yao and Fang Dai and Peng Sun and Weituo Zhang and Biyun Qian and Hui Lu : Enhancing the fairness of AI prediction models by Quasi-Pareto improvement among heterogeneous thyroid nodule population, Nature Communications, vol.15, p.1958 (2024)

\bibitem{gessert2020skin}
Gessert, Nils and Nielsen, Maximilian and Shaikh, Mohsin and Werner, Ren{\'e} and Schlaefer, Alexander: Skin lesion classification using ensembles of multi-resolution EfficientNets with meta data, MethodsX, vol.7, p.100864, (2020)

\bibitem{plataniotis2000color}
Plataniotis, Konstantinos N. and Venetsanopoulos, Anastasios N.:Color Image Processing and Applications, Springer-Verlag, (2000)

\bibitem{basu2002gaussian}
Basu, Mitra: Gaussian-based edge-detection methods—A survey, IEEE Transactions on Systems, Man, and Cybernetics, Part C (Applications and Reviews), vol.32, pp.252--260, (2002)

\bibitem{kong2024achieving}
Kong, Qingpeng and Chiu, Ching-Hao and Zeng, Dewen and Chen, Yu-Jen and Ho, Tsung-Yi and Hu, Jingtong and Shi, Yiyu:Achieving Fairness Through Channel Pruning for Dermatological Disease Diagnosis, arXiv preprint arXiv:2401.12345, (2024)

\bibitem{zhou2016cam}
Zhou, Bolei and Khosla, Aditya and Lapedriza, Agata and Oliva, Aude and Torralba, Antonio: Learning Deep Features for Discriminative Localization, Proceedings of the IEEE Conference on Computer Vision and Pattern Recognition (CVPR), pp.2921--2929, (2016)

\bibitem{selvaraju2017grad}
Selvaraju, Ramprasaath R and Cogswell, Michael and Das, Abhishek and Vedantam, Ramakrishna and Parikh, Devi and Batra, Dhruv: Grad-CAM: Visual Explanations from Deep Networks via Gradient-based Localization, Proceedings of the IEEE International Conference on Computer Vision (ICCV), pp.618--626 (2017)

\bibitem{quadrianto2019discovering}
Quadrianto, Novi and Sharmanska, Viktoriia and Thomas, Oliver: Discovering Fair Representations in the Data Domain, In: Proceedings of the IEEE/CVF conference on computer vision and pattern recognition. pp. 8227–8236 (2019)

\bibitem{jung2021fair}
Jung, Sangwon and Lee, Donggyu and Park, Taecon and Moon, Taesup: Fair Feature Distillation for Visual Recognition,  In: Proceedings of the IEEE/CVF conference on computer vision and pattern
 recognition. pp. 12115–12124 (2021)

\bibitem{xu2023fairadabn}
Xu, Z., Zhao, S., Quan, Q., Yao, Q., Zhou, S.K.: Fairadabn: Mitigating unfairness with adaptive batch normalization and its application to dermatological disease classification. arXiv preprint arXiv:2303.08325 (2023)

\bibitem{xu2025incorporating}
Xu, Gelei and Duan, Yuying and Liu, Zheyuan and Li, Xueyang and Jiang, Meng and Lemmon, Michael and Jin, Wei and Shi, Yiyu: Incorporating Rather Than Eliminating: Achieving Fairness for Skin Disease Diagnosis Through Group-Specific Experts, Medical Image Computing and Computer Assisted Intervention -- MICCAI 2025, (2025)

\bibitem{wu2023fairprune}
Wu, Yawen and Zeng, Dewen and Xu, Xiaowei and Shi, Yiyu and Hu, Jingtong:FairPrune: Achieving Fairness Through Pruning for Dermatological Disease Diagnosis,Proceedings of the IEEE/CVF Conference on Computer Vision and Pattern Recognition (CVPR), pp.743--753, (2023)

\bibitem{corbin2023assessing}
Corbin, Adam and Marques, Oge: Assessing Bias in Skin Lesion Classifiers With Contemporary Deep Learning and Post-Hoc Explainability Techniques, IEEE Access, vol.11, pp.78341--78352, (2023)

\bibitem{he2016deep}
He, Kaiming and Zhang, Xiangyu and Ren, Shaoqing and Sun, Jian: Deep Residual Learning for Image Recognition,Proceedings of the IEEE Conference on Computer Vision and Pattern Recognition (CVPR),pp.770--778, (2016)

\bibitem{ronneberger2015unet}
Ronneberger, Olaf and Fischer, Philipp and Brox, Thomas: U-Net: Convolutional Networks for Biomedical Image Segmentation, arXiv preprint arXiv:1505.04597, (2015)

\bibitem{biaslessnas}
Yi Sheng, Junhuan Yang, Jinyang Li, James Alaina, Xiaowei Xu, Yiyu
 Shi, Jingtong Hu, Weiwen Jiang, and Lei Yang: Data-Algorithm-Architecture Co-Optimization for Fairness in Dermatology, Medical Image Computing and Computer Assisted Intervention -- MICCAI, pp.1--11, (2023)
 
\bibitem{guo2023calibrating}
Guo, Cheng-Yan and Huang, Wen-Yao and Chang, Hao-Ching and Hsieh, Tung-Li: Calibrating Oxygen Saturation Measurements for Different Skin Colors Using the Individual Typology Angle, IEEE Sensors Journal, vol.23, pp.16993-17001, (2023)

\bibitem{el-khatib2020deep}
El-Khatib, Hassan and Popescu, Dan and Ichim, Loretta: Deep Learning-Based Methods for Automatic Diagnosis of Skin Lesions, Sensors, vol.20, p.1753, (2020)

\bibitem{pundhir2023ethical}
Pundhir, Anshul and Verma, Sanchit and Raman, Balasubramanian: Towards Ethical Dermatology: Mitigating Bias in Skin Condition Classification, arXiv preprint arXiv:2305.12345, (2023)

\bibitem{mustafa2025deep}
Mustafa, Saleem and Jaffar, Arfan and Rashid, Muhammad and Akram, Sheeraz and Bhatti, Sohail Masood: Deep learning-based skin lesion analysis using hybrid ResUNet++ and modified AlexNet-Random Forest for enhanced segmentation and classification, PLOS ONE, vol.20, p.1, (2025)

\bibitem{karras2019stylebased}
Karras, Tero and Laine, Samuli and Aila, Timo: A Style-Based Generator Architecture for Generative Adversarial Networks, Proceedings of the IEEE/CVF Conference on Computer Vision and Pattern Recognition (CVPR), pp.4401--4410, (2019)

\bibitem{Chen2024SkinCancer}
Jennifer Y. Chen and Kristen Fernandez and Raj P. Fadadu and Rasika Reddy and Mi-Ok Kim and Josephine Tan and Maria L. Wei, Skin Cancer Diagnosis by Lesion, Physician, and Examination Type: A Systematic Review and Meta-Analysis, JAMA Dermatology, vol.161, pp.135--146, (2025)

\bibitem{Paxton2024MeasuringAIFairness}
Kuniko Paxton and Koorosh Aslansefat and Dhavalkumar Thakker and Yiannis Papadopoulos: Measuring AI Fairness in a Continuum Maintaining Nuances: A Robustness Case Study, vol.28, pp.17--19, (2024)

\bibitem{adadi2018peeking}
Adadi, Amina and Berrada, Mohammed: Peeking Inside the Black-Box: A Survey on Explainable Artificial Intelligence (XAI): IEEE Access, vol.6, pp.52138--52160, (2018)

\bibitem{sharma2004ciede2000}
Sharma, Gaurav and Wu, Wencheng and Dalal, Edul N.: The {CIEDE2000} Color-Difference Formula: Implementation Notes, Supplementary Test Data, and Mathematical Observations: Color Research and Application, vol.30, pp.21--30, (2005)

\bibitem{hardt2016equality}
Hardt, Moritz and Price, Eric and Srebro, Nati: Equality of Opportunity in Supervised Learning, arXiv preprint arXiv:1610.02413, (2016)

\bibitem{borchers2025abroca}
Borchers, C., \& Baker, R. S. (2025). ABROCA Distributions For Algorithmic Bias Assessment: Considerations Around Interpretation. The 15th International Learning Analytics and Knowledge Conference (LAK 2025),(2025)

\bibitem{ansari2024algorithmic}
Ansari, Faizanuddin and Chakraborti, Tapabrata and Das, Swagatam:Algorithmic Fairness in Lesion Classification by Mitigating Class Imbalance and Skin Tone Bias, Medical Image Computing and Computer Assisted Intervention – MICCAI 2024, pp.375--382, (2024)

\bibitem{bhattacharyya2024conformal}
Bhattacharyya, Swarnava and Pal, Umapada and Chakraborti, Tapabrata:Conformal uncertainty quantification to evaluate predictive fairness of foundation AI model for skin lesion classes across patient demographics, arXiv:2503.23819v1, (2024)

\bibitem{chowdhury2021exploring}
Chowdhury, Tamal and Bajwa, Angad R.S. and Chakraborti, Tapabrata and Rittscher, Jens and Pal, Umapada:Exploring the Correlation Between Deep Learned and Clinical Features in Melanoma Detection, Medical Image Understanding and Analysis, pp.1--15, (2021)

\bibitem{ren2024skincon}
Ren, Zhihang and Li, Yunqi and Li, Xinyu and Xie, Xinrong and Duhaime, Erik P. and Fang, Kathy and Chakraborti, Tapabrata and Guo, Yunhui and Yu, Stella X. and Whitney, David:SkinCON: Towards Consensus for the Uncertainty of Skin Cancer Sub-typing Through Distribution Regularized Adaptive Predictive Sets (DRAPS), Medical Image Computing and Computer Assisted Intervention--MICCAI 2024: 27th International Conference, Proceedings, Part I, pp.405--415, (2024)
\end{thebibliography}
\end{document}